\documentclass{aastex}

\usepackage{graphicx}

\usepackage{emulateapj5}
\usepackage{times}

\makeatletter

\newenvironment{inlinefigure}{%
\def\@captype{figure}%
\noindent\begin{minipage}{0.999\linewidth}\begin{center}}
{\end{center}\end{minipage}\smallskip}
\makeatother

\usepackage{mathptmx}

\def\gax    {{_>\atop^{\sim}}}
\def\Mo     {{\rm M}_{\odot}}
\def\Lo     {{\rm L}_{\odot}}

\begin{document}

\slugcomment{submitted to {\em The Astrophysical Journal}}

\title{The Diffuse Emission and a Variable ULX in the Elliptical Galaxy NGC 3379}

\author{Laurence P. David, Christine Jones, William Forman \& Steve Murray}
\affil{Harvard-Smithsonian Center for Astrophysics, 60 Garden St.,
Cambridge, MA 02138;\\ david@cfa.harvard.edu}

\shorttitle{\emph ULX in NGC3379}
\shortauthors{David, Jones, Forman \& Murray}

\begin{abstract}

A Chandra observation of the intermediate luminosity ($M_B=-20$) elliptical 
galaxy NGC 3379 resolves 75\% of the X-ray emission within the central 5~kpc into point sources.  
Spectral analysis of the remaining unresolved emission within the central 770~pc indicates
that 90\% of the emission probably arises from undetected point sources, while 10\% arises
from thermal emission from $kT=0.6$~keV gas. Assuming a uniform density distribution
in the central region of the galaxy gives a gas mass of $5 \times 10^5 \Mo$.  
Such a small amount of gas can be supplied by stellar mass loss in only $10^7$~years.  Thus, 
the gas must be accreting into the central supermassive
black hole at a very low radiative efficiency as in the ADAF or RIAF models, or is being 
expelled in a galactic wind driven by the same AGN feedback mechanism as that 
observed in cluster cooling flows.  If the gas is being expelled in 
an AGN driven wind, then the ratio of mechanical to radio power of the AGN must 
be $10^4$, which is comparable to that measured in cluster cooling flows which have recently 
been perturbed by radio outbursts.  Only 8\% of the detected point sources are coincident
with globular cluster positions, which is significantly less than that found among
other ellipticals observed by Chandra.  The low specific frequency of globular clusters
and the small fraction of X-ray point sources associated with globulars in NGC 3379 is more 
similar to the properties of lenticular galaxies rather than ellipticals.

The brightest point source in NGC 3379 is located 360~pc from the central AGN 
with a peak luminosity of $3.5 \times 10^{39}$ergs~s$^{-1}$, which places it in the 
class of ultra-luminous X-ray point sources (ULX).  Analysis of an archival ROSAT HRI observation
of NGC 3379 shows that this source was at a comparable luminosity 5 years prior to the
Chandra observation.  The spectrum of the ULX is well described by a power-law model
with $\Gamma=1.6 \pm 0.3$ and galactic absorption, similar to other ULXs observed
by Chandra and XMM-Newton and to the low-hard state
observed in galactic black hole binaries.  During the Chandra observation, the source 
intensity smoothly varies by a factor of two with the suggestion of an 8-10 hour period.  
No changes in hardness ratio are detected as the intensity of the source varies.
While periodic behavior has recently been detected in several ULXs, all of these reside
within spiral galaxies.  The ULX in NGC 3379 is the only known ULX in an elliptical
galaxy with a smoothly varying light curve suggestive of an eclipsing binary system.
  
\end{abstract}


\section{Introduction}

Einstein observations showed that the bulk of the X-ray 
emission from optically luminous early-type galaxies arises
from hot gas in hydrostatic equilibrium (Forman, Jones \& Tucker 1985). 
A spectrally harder X-ray component, more prevalent among low luminosity 
ellipticals, was detected by ASCA and assumed to arise from 
low-mass X-ray binaries (LMXBs) due to the old stellar population in
these galaxies (Kim, Fabbiano \& Trinchieri 1992).  
Chandra, with its superior angular resolution, has resolved populations of 
point sources in many early-type galaxies (e.g., Angelini, Lowenstein \& Mushotzky 2001;
Sarazin, Irwin \& Bregman 2001; Kraft et al. 2001; Blanton, Sarazin \& Irwin 2001;
Finogueov \& Jones 2002, Jeltema et al. 2003). 
Between 20 and 80\% of the point sources detected in early-type galaxies reside in 
globular clusters (Sarazin et al. 2003).
This is consistent with observations of our own galaxy which show that LMXBs 
are preferentially located in globular clusters compared to the rest of the galaxy
(White, Nagase \& Van den Heuvel 1995).

Ultra-luminous X-ray point sources (ULXs), 
off-nuclear point sources with $L_x \gax 10^{39}$ergs~s$^{-1}$,
have been reported in approximately 15 early-type galaxies (e.g., Angeline et al. 2001;
Sarazin, Irwin \& Bregman 2001; Jeltema et al. 2003; Swartz et al. 2004).
However, Irwin, Bregman \& Athey (2004) argue that the statistics of ULXs
in early-type galaxies with $L_x > 2 \times 10^{39}$ergs~s$^{-1}$
are consistent with background sources.  Swartz et al. (2004) analyzed archival Chandra 
observations of 82 galaxies and compiled a sample of 154 ULXs.  They list 
two ULXs in NGC 3379 and also report that the the most luminous source is likely variable. 
The minimum luminosity for
an ULX corresponds to the Eddington luminosity of the most massive 
black holes expected to form from stellar core collapse (Fryer \& Kalogera 2001). 
The nature and origin of ULXs are unknown, but they could arise from 
non-isotropic emission from stellar mass black holes (King et al. 2001),
super-Eddington emission from stellar mass black holes with inhomogeneous disks (Begelman 2002),
or sub-Eddington emission from intermediate mass black holes (IMBHs).

NGC 3379 is an intermediate luminosity ($M_B=-20.0$) E1 galaxy in the Leo I group
at a distance of 10.6~Mpc (Tonry et al. 2001).  The large scale optical morphology of NCG3379
is nearly featureless and well fitted with a de Vaucouleurs profile (Capaccioli et al. 1990).
Terlevich \& Forbes (2002) estimate a stellar age of 9.3 Gyr for NGC 3379.
Both of these characteristics indicate that NGC 3379 has not undergone a major merger in the recent 
past.  However, warm ionized gas and dust have been detected in the central few arcseconds of NGC 3379

\begin{inlinefigure}
  \center{\includegraphics*[width=1.00\linewidth,bb=60 208 470 562,clip]{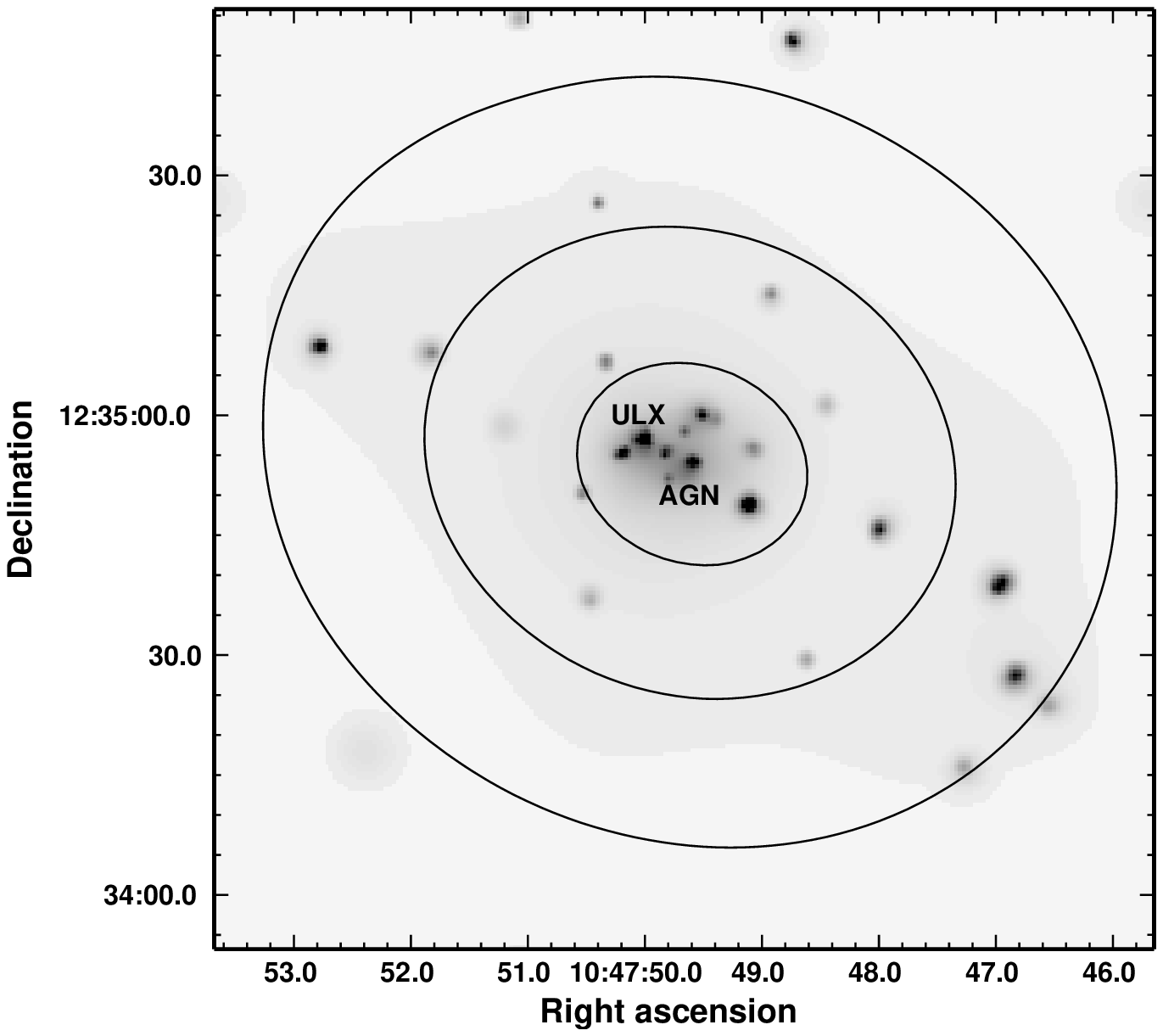}}
  \caption{Adaptively smoothed 0.3-6.0~keV ACIS-S3 image of the central 
  $2^{\prime}$ by $2^{\prime}$ (6.2~kpc) region of NGC 3379 along with 
  the optical isophotes of the galaxy.  The AGN at the galactic
  center and the ULX are labeled.}
\end{inlinefigure}

\bigskip

\noindent
(Macchetto et al. 1996, van Dokkum \& Franx 1995).
Statler (2001) argues that the $1.5^{\prime\prime}$ (77~pc) dust ring is dynamically decoupled from the 
stars and likely has an external origin.
There is also some evidence for a central stellar disk based on the stellar
rotation curves (Pastoriza et al. 2000) along with a central 
compact object with a mass of $1.0-3.9 \times 10^8~ \Mo$
(Magorrian et al. 1998, Haring \& Rix 2004).
Thus, while the outer regions of NGC 3379 are dynamically relaxed, there are 
some indications of recent merger activity in the central parts of the galaxy.

In $\S 2$ we discuss the general properties of the X-ray point
population detected in the Chandra observation of NGC 3379, including the luminosity
function, spatial distribution and hardness ratios.  In $\S 3$ we discuss the spectral and temporal 
characteristics of the ULX.

\noindent
The properties of the diffuse emission and the central AGN 
are presented in $\S 4$. A discussion concerning the nature of the ULX and 
the dynamic state of the hot gas in NGC 3379 is given in $\S 5$, followed
by a brief summary of our main results in $\S 6$.

\section{The Point Source Population}

NGC 3379 was observed by Chandra for 33,744~s on Feb. 13, 2001 with the ACIS-S detector
in faint telemetry mode.  The data were reprocessed with the CIAO 3.2 version of 
{\it acis\_process\_events} along with CALDB 3.0. Filtering the S3 data for background
flares leaves 31,199~s of cleaned data. Since the archived data were processed several
years ago, we also followed the {\it fix off-sets} thread on the CXC web pages to improve 
the astrometry, but this only produced a shift of $0.12^{\prime\prime}$.  
We then checked for X-ray detections of stars in the USNO-B1.0 catalog and found
one star with an optical position off-set by $0.2^{\prime\prime}$ from its X-ray position,
which is within the absolute astrometry uncertainties of Chandra and the USNO B1 catalog.

An adaptively smoothed 0.3-6.0~keV S3 image of the central $2^{\prime}$
by $2^{\prime}$ region (6.2~kpc on a side)
is shown in Fig. 1 along with the optical isophotes of the galaxy.  
This image shows that most of the X-ray emission from NGC 3379 is resolved by Chandra 
into point sources, many of which are 
aligned along the major optical axis of the galaxy.  
The point source labeled AGN in Fig.

\begin{inlinefigure}
  \center{\includegraphics*[width=1.00\linewidth,bb=135 267 414 506,clip]{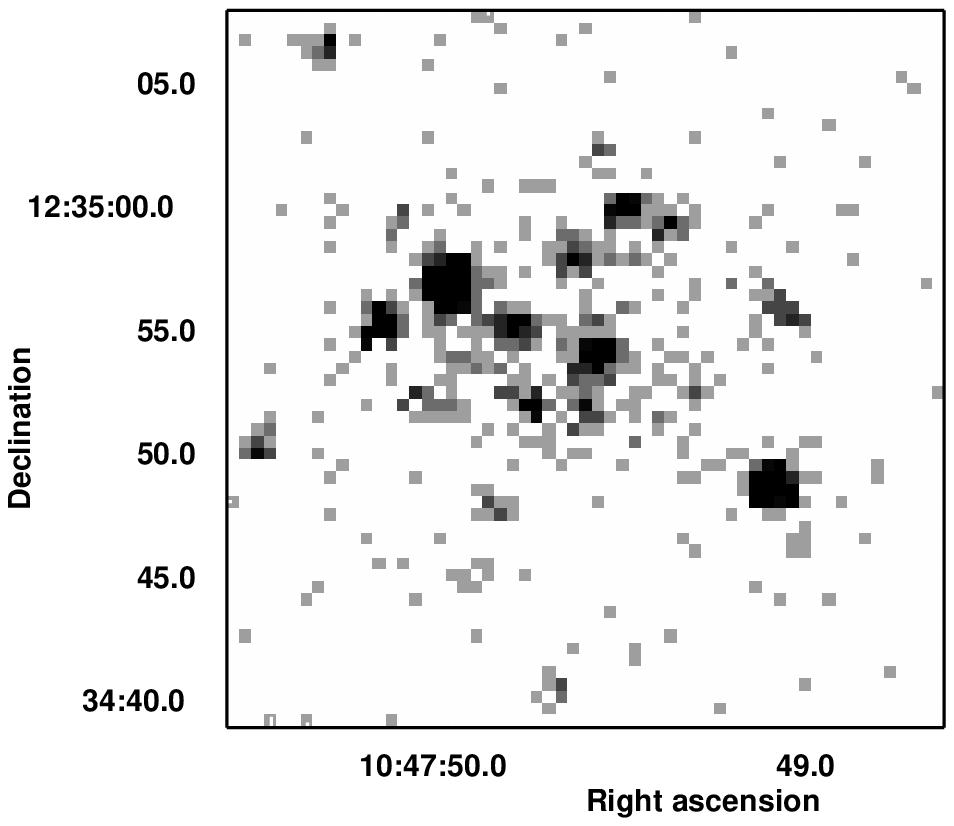}}
  \caption{Raw 0.3-6.0~keV ACIS-S3 image of the central $30^{\prime\prime}$ by $30^{\prime\prime}$ 
  (1.5~kpc) region of NGC 3379.}
\end{inlinefigure}

\bigskip

\noindent
1 is located within $1^{\prime\prime}$ 
of the optical centroid of the galaxy, and is probably associated with the central 
supermassive black hole with a dynamically measured mass of $1.0-3.9 \times 10^8 \Mo$
(Magorrian et al. 1998, Haring \& Rix 2004).
The brightest source in NGC 3379 (labeled ULX in Fig. 1) is located $7^{\prime\prime}$ (360 pc)
to the NE of the AGN.  
The center of NGC 3379 has a high surface density of point sources,
as can be seen in the full resolution, raw data image of the central 
$30^{\prime\prime}$ by $30^{\prime\prime}$ (1.5~ kpc) region shown in Fig. 2.

Using a wavelet detection algorithm on the 0.3-6.0~keV image with a detection
threshold of $10^{-6}$, we detect 66 sources within the S3 field of view.  
The point source detection sensitivity varies across the field of view
due to the presence of extended emission in the center of the galaxy
with an average, exposure-corrected $3 \sigma$ threshold of $3.1 \times 10^{-4}$~ct~s$^{-1}$.  
Assuming an absorbed power-law model with galactic absorption 
($N_H = 2.79 \times 10^{20}$~cm$^{-2}$) and 
index $\Gamma=1.7$, this count rate corresponds to an unabsorbed 0.3-10.0~keV flux of 
$1.6 \times 10^{-15}$~ergs~cm$^{-2}$~s$^{-1}$ and luminosity 
L(0.3-10.0~keV) = $2.2 \times 10^{37}$ergs~s$^{-1}$.
Forty sources above this threshold are detected in the central $1.6^{\prime}$ (5~kpc).
Based on the Chandra deep field south (Giacconi et al. 2001), approximately one
of these sources is likely a background object.  Beyond 5~kpc, 
approximately 7 of the detected 26 sources are likely background objects.  
We therefore restrict most analysis to the central 5 kpc of the galaxy.

Based on our count rate conversion factor, 
the two brightest sources in NGC 3379 have luminosities of 
$2.4 \times 10^{39}$~ergs~s$^{-1}$ and $6.8 \times 10^{38}$~ergs~s$^{-1}$
The luminosity of the brightest source is consistent with that obtained 
by Swartz et al. (2004), but the luminosity of the second brightest
source is a factor of 1.8 less than that given by Swartz et al., after
correcting for the slightly larger distance used by Swartz et al. 
of 11.1~Mpc from Ferrarese et al. (2000).
Swartz et al. derived unabsorbed X-ray fluxes in the 0.5-8.0~keV band pass
by fitting an absorbed power-law model to the spectra of each candidate
ULX, treating the absorption and power-law index as free parameters.
They derived best-fit parameters of $N_H=2.2 \times 10^{21}$~cm$^{-2}$ and
$\Gamma=2.58$ for the second brightest 
source in NGC 3379, which accounts for

\begin{inlinefigure}
  \center{\includegraphics*[width=0.90\linewidth,bb=20 150 565 700,clip]{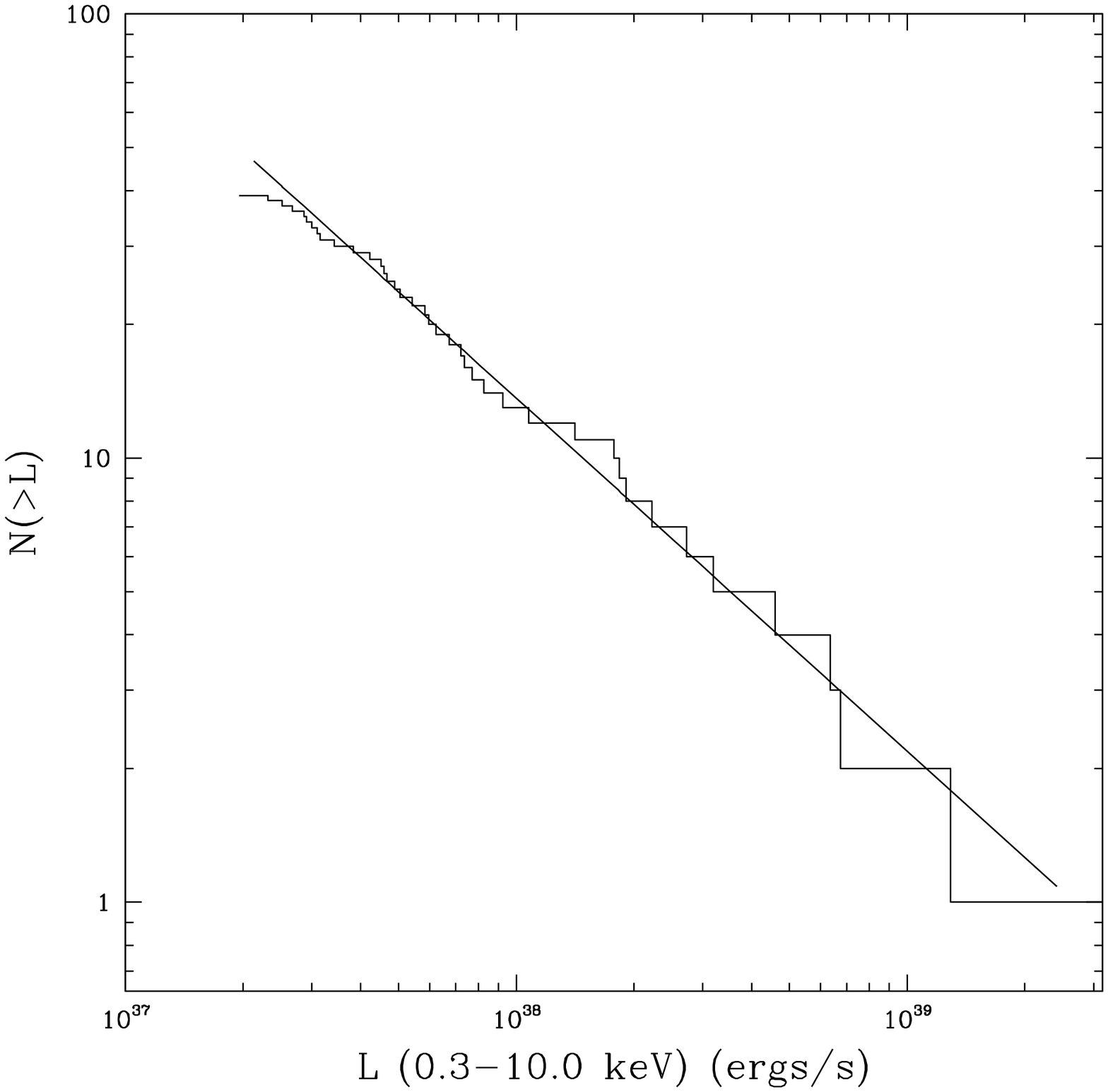}}
  \caption{Cumulative luminosity function of all point sources detected within 
  the central $5^{\prime}$, excluding the central AGN.  Also shown is the
  best-fit power law luminosity function for sources detected at greater
  than $4 \sigma$}
\end{inlinefigure}

\bigskip

\noindent
much of the difference between
our two luminosity estimates for the second brightest source in NGC 3379.

\subsection{Luminosity Function and Spatial Distribution}

Excluding the central AGN, there are 39 detected sources in the central 
5~kpc of the galaxy (see the luminosity function in Fig. 3). We first
fitted the luminosity function of all 39 sources using the maximum likelihood
method in Crawford, Jauncey \& Murdoch (1970) to a power-law model 
($N(>L)=k L^{-\alpha}$) and obtained a best-fit
index of $\alpha=0.60\pm 0.15$ ($1 \sigma$ error). This gives a reasonable 
fit at low fluxes, but significantly overestimates the number of 
brighter sources.  We then repeated the analysis only including
sources detected at more than $4 \sigma$ and obtained
an acceptable fit with $\alpha=0.80\pm 0.2$ (see Fig. 3).
The luminosity function does not have a
break near the Eddington-limit of a neutron star 
($L_{edd} = 1.6 \times 10^{-15}$~ergs~cm$^{-2}$~s$^{-1}$) as observed
in some early-type galaxies (Sarazin et al. 2001). 
The flattening of the luminosity function at low fluxes may be due to the presence
of extended emission in the center of the galaxy which raises the detection
threshold in this region.

Kim \& Fabbiano (2004) recently derived the luminosity function for a sample
of early-type galaxies, including NGC 3379, but they only included sources
beyond the central $20^{\prime\prime}$ and within the $D_{25}$ ellipse
($5.4^{\prime}$ by $4.8^{\prime}$).  
However, they obtained a best-fit index of $\beta=1.8$ for the differential
luminosity function, which corresponds to the 
same index we find for the cumulative luminosity function.
Colbert et al. (2004) calculated the point source luminosity function 
in 23 late-type galaxies and 9 early-type galaxies, including NGC 3379.
Fitting the luminosity function of the 12 sources with luminosities greater 
than $10^{38}$~ergs~s$^{-1}$ and within the $D_{25}$ ellipse, they obtained a 
best-fit index for the cumulative luminosity function of $1.07$, which 
is consistent with our results within the errors. 
In general, star forming galaxies have flatter luminosity functions
than early-type galaxies.  For example, Colbert et al. find an average index for 
spiral and starburst galaxies of of $\alpha=0.6-0.8$ compared to 
ellipticals with $\alpha \approx 1.4$.  However, the average slope
of the 

\begin{inlinefigure}
  \center{\includegraphics*[width=0.90\linewidth,bb=20 150 565 700,clip]{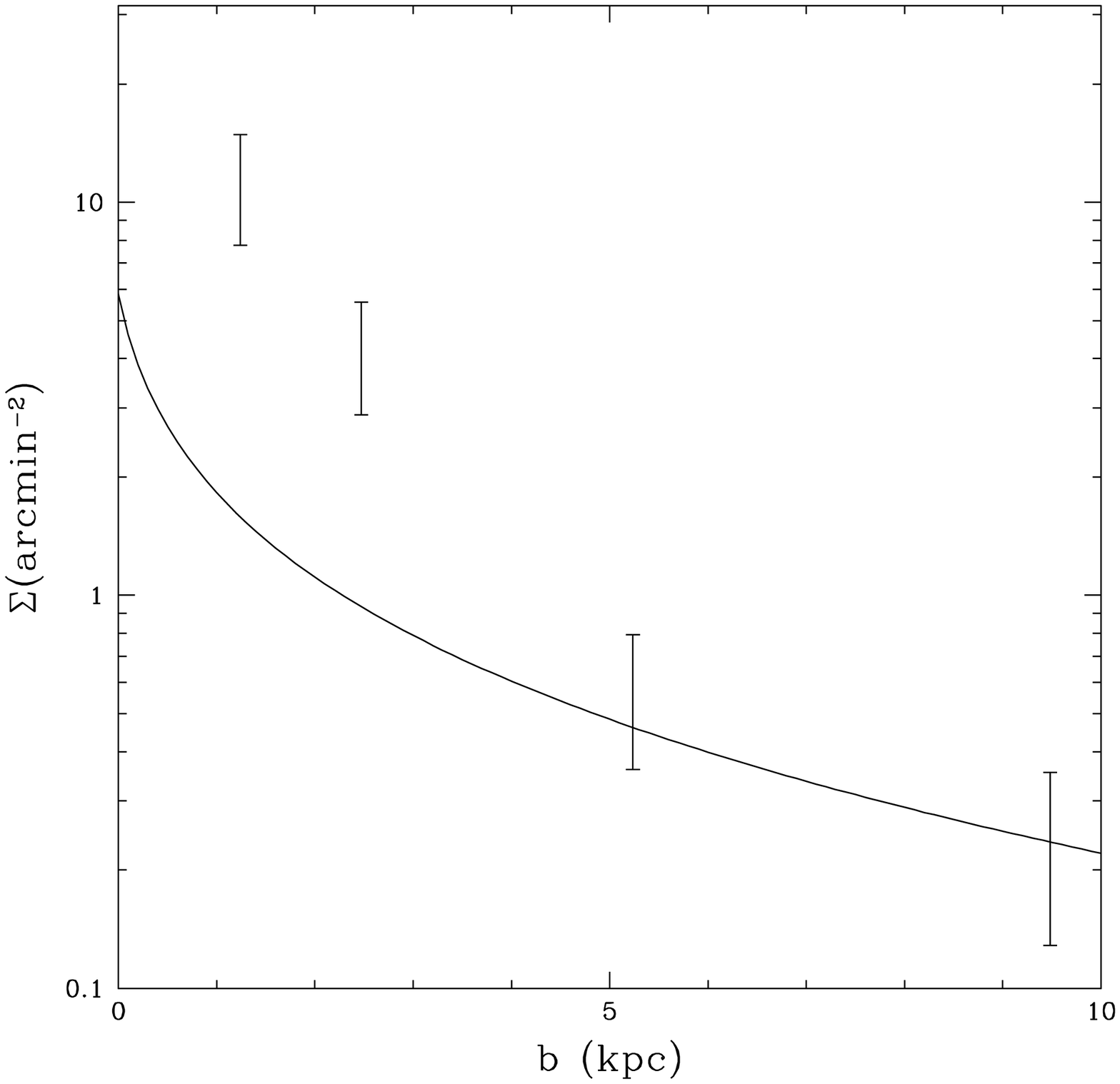}}
  \caption{Surface brightness profile of the detected X-ray point sources in 
  NGC 3379 along with the optical surface brightness profile (solid line).
  The normalization of the optical surface has been adjusted for comparison.}
\end{inlinefigure}

\bigskip

\noindent
14 ellipticals analyzed in the Kim \& Fabbiano (2004) sample
is $\alpha=0.9$, which is consistent with our result for NGC 3379.

While the ratio of the combined X-ray luminosity of the detected point sources in 
NGC 3379 to the optical luminosity of the galaxy is consistent with other early-type 
galaxies observed by Chandra (Colbert et al. 2004),  the point sources
have a much more centrally peaked surface brightness profile
than the optical light of the galaxy (see Fig. 4).  The optical surface brightness 
profile in Fig. 4 is based on the de Vaucouleurs profile for NGC 3379 found by 
Capaccioli et al. (1990) with $r_e=54.8^{\prime\prime}$ and 
$\mu_e = 22.24$~mag~arcsec$^{-2}$.  For comparison, Finoguenov \& Jones (2002) found
that there was a deficit of detected point sources in the core of M84 compared to the
optical surface brightness profile of the galaxy.  However, there is significantly more
diffuse emission in M84 compared to NGC 3379, and the deficit of point sources in 
the core may be due to the lower sensitivity of detecting point sources in the presence
of diffuse emission. Accounting for the lower sensitivity of detecting point sources in 
the core of NGC 3379, would only increase the excess of point sources within the central few kpc.

\bigskip

\subsection{Hardness Ratios}

Of all the discrete sources in NGC 3379, 
only the ULX has sufficient counts for a detailed spectral
analysis (see $\S 3$).  To help in the identification of the detected point
sources, we followed Prestwich et al. (2003) and extracted counts in a
soft band from 0.3-1.0~keV, a medium band from 1.0-2.0~keV and a hard
band from 2.0-8.0~keV.  We then computed a soft X-ray color from (M-S)/T
and a hard X-ray color from (H-M)/T, where T is the 
total number of counts in all 3 bands.  The resulting color-color
diagram for all non-nuclear sources detected at more than $4 \sigma$ and within the central 5~kpc
is shown in Fig. 5.  The circle in Fig. 5 delineates the color-color region associated 
with LMXBs determined by Prestwich et al.  This figure shows that all
of the sources detected in NGC 3379 have X-ray colors consistent with LMXBs.
Thermal supernova remnants and 

\begin{inlinefigure}
  \center{\includegraphics*[width=0.90\linewidth,bb=20 150 565 700,clip]{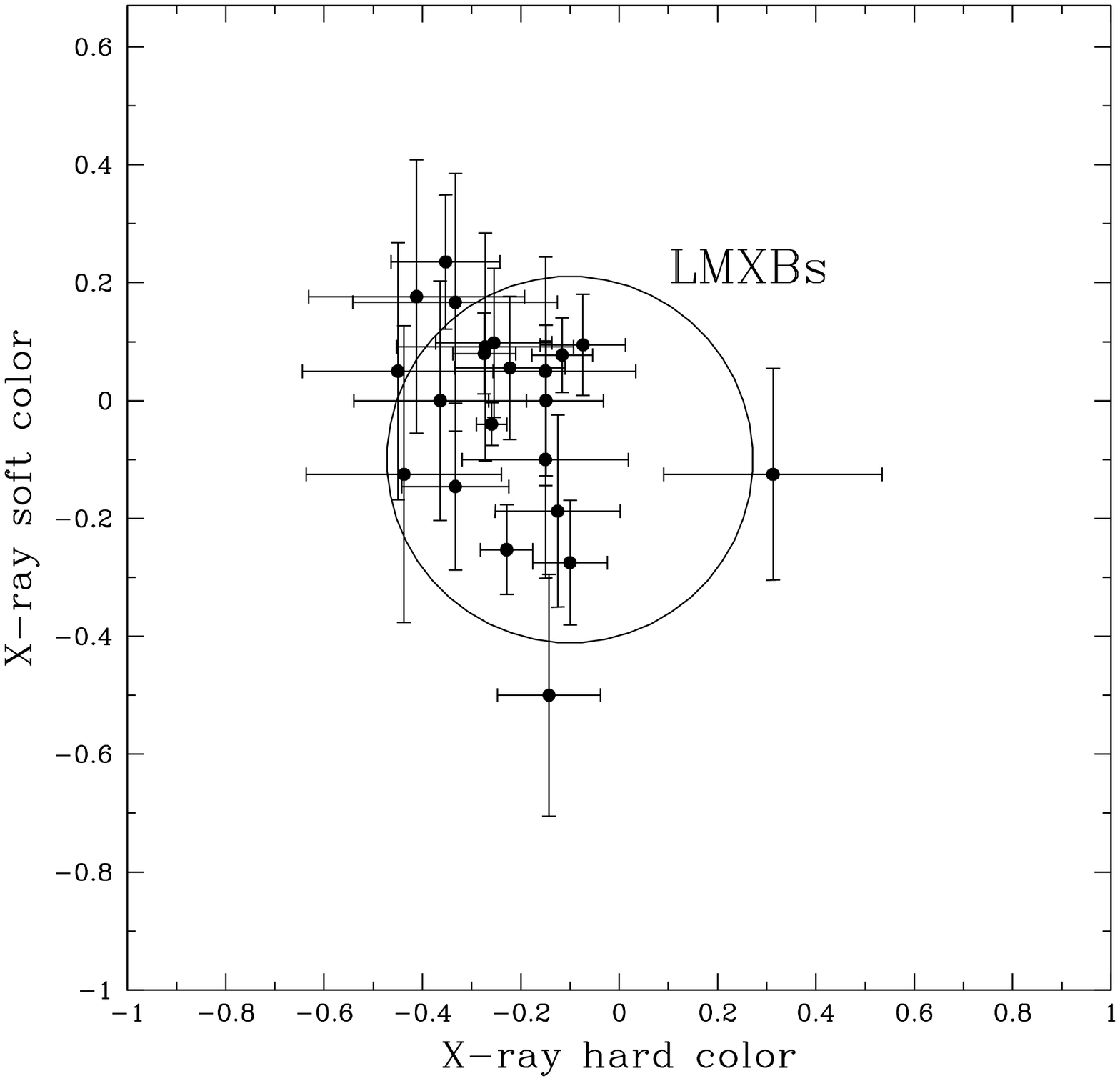}}
  \caption{Color-color diagram for all non-nuclear point sources detected at more than $4 \sigma$ 
  significance within the central 5 kpc. The circle shows the
  range of colors of LMXBs identified in Local Group galaxies.}
\end{inlinefigure}

\bigskip

\noindent
super-soft sources have softer colors,
while HMXBs typically have harder colors.

\subsection{Fraction in Globular Clusters}

Whitlock, Forbes \& Beasley (2003) list the positions of 133 globular clusters in NGC 3379,
of which 105 are within the S3 field-of-view.   
They find that the net surface brightness of globular clusters is consistent
with zero beyond $5.5^{\prime}$ from the center of the galaxy which roughly 
corresponds to the farthest corner of the S3 chip.  Of the 65 non-nuclear point sources 
detected in the S3 image, only 3 are within $2^{\prime\prime}$ of the globular 
cluster positions in Whitlock et al.  Fig. 6 shows shows the raw 0.3-6.0~keV S3 data for the 
central 10~kpc by 10~kpc region of NGC 3379, along with $2^{\prime\prime}$ radius circles at the 
globular cluster positions.  Within the central 5~kpc, only 2 (identified as S1 and S2 in 
Fig. 6), of the 39 non-nuclear point sources are coincident with globular clusters.
This figure shows that the globular cluster sample is highly incomplete within the 
central $20^{\prime\prime}$ due to the difficulty of detecting globulars against the 
brightest parts of the galaxy.  However, even if we 
exclude the central $20^{\prime\prime}$ and consider only the remaining 25 point sources 
within the central 5~kpc, only 2 (8\%) of these sources are identified with globular clusters.
This is significantly less than the 50\% typically
found in early-type galaxies (Sarazin et al. 2003).  The X-ray luminosity threshold 
in our observation is $2.2 \times 10^{37}$ergs~s$^{-1}$, which is 
fainter than that in most other studies of the binary populations in early-type
galaxies, so incompleteness in the X-ray point source sample should not be a factor.
Kundu \& Whitmore (2001) find that the turnover absolute magnitude of the 
globular cluster luminosity function for a sample of early-type galaxies, including NGC3379,
is $M_v=-7.41$.  The lower magnitude cut in the Whitlock et al. globular cluster
sample is $m_B=23$, corresponding to $m_v \approx 22.3$ for NGC 3379. Using the distance modulus 
for NGC 3379, the turnover magnitude in the globular cluster luminosity function should be 
$m_v^0 \approx 22.7$, which is slightly fainter than the magnitude cut in the Whitlock et al.
sample.  Thus, incompleteness in the globular cluster sample

\begin{inlinefigure}
  \center{\includegraphics*[width=1.00\linewidth,bb=42 182 486 578,clip]{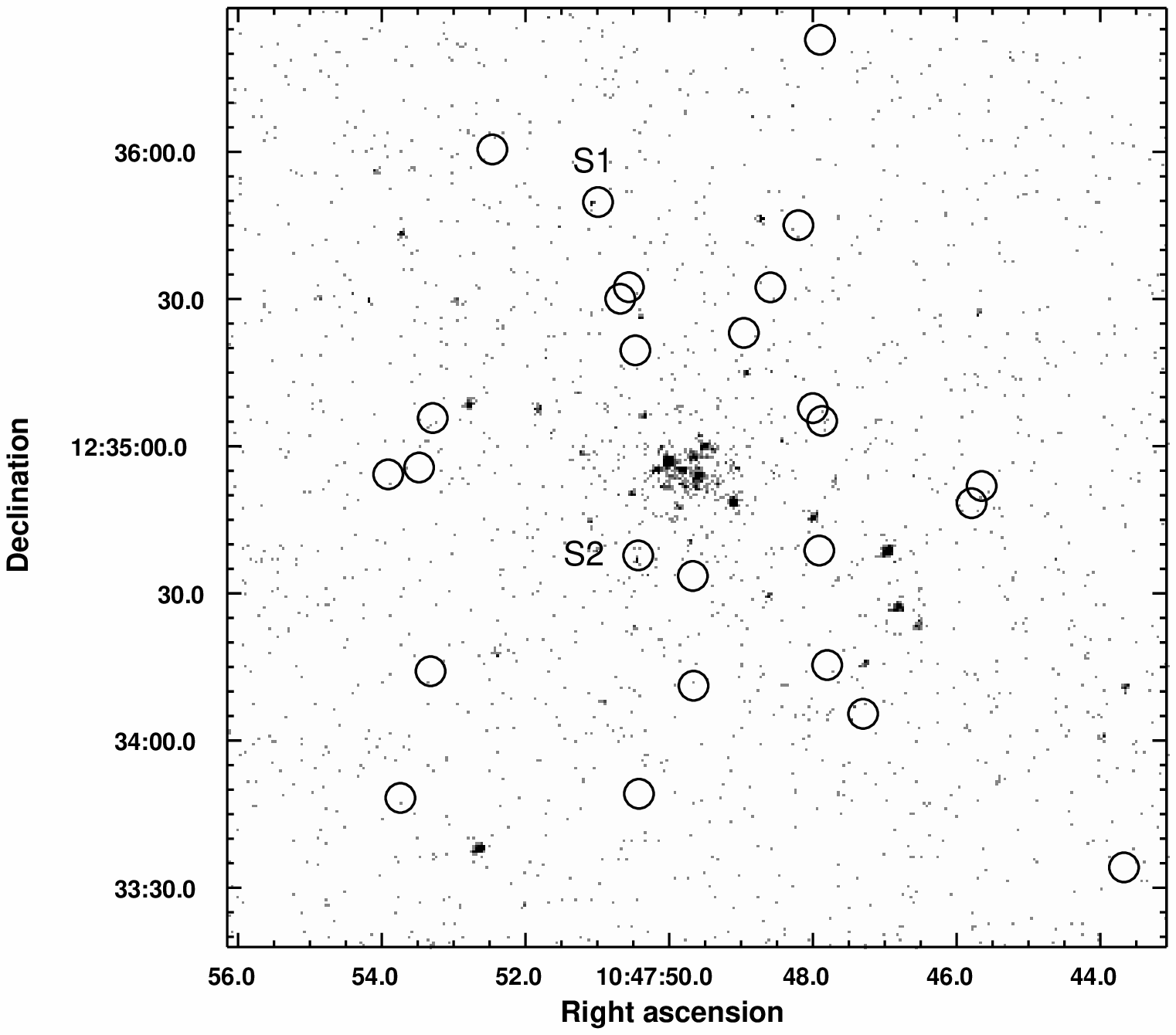}}
  \caption{The raw 0.3-6.0~kev S3 image of NGC3379.  The image is 10~kpc on a side.
  Also shown are the locations of the globular clusters from 
  Whitlock, Forbes \& Beasley (2003). Only two X-ray point sources, labeled S1 and S2,
  are within $2^{\prime\prime}$ of the globular cluster positions.}
\end{inlinefigure}

\bigskip

\noindent
should not have a significant impact on the small fraction of X-ray point sources associated 
with globular clusters in NGC 3379, especially since Sarazin et al. (2003) find that
X-ray point sources in early-type galaxies tend to be found in the optically more 
luminous globular clusters.

NGC3379 has a very low specific frequency of globular clusters for an elliptical
galaxy, $S_N=1.2$ (Kundu \& Whitmore 2001), which is equal to the mean found for
a sample of Sa and Sb galaxies (Harris 1991). 
There is some evidence that the fraction of X-ray point sources associated with globular clusters
increases from late to early-type galaxies (Sarazin et al. 2003).  
The fraction in NGC 3379 is actually consistent with our Galaxy, within
which approximately 10\% of LMXBs are associated with globular clusters 
(White, Nagase \& Van den Heuvel 1995).
It is interesting to note that there has been significant debate over the
years whether NGC 3379 is an elliptical or a low inclination S0 galaxy
as suggested by Capaccioli et al. (1990).  The low specific frequency 
of globular clusters in NGC 3379 and the small fraction of X-ray point sources
associated with globular clusters are consistent with the Chandra observation
of the S0 galaxy NGC 1553 (Blanton, Sarazin \& Irwin 2001).

\section{ULX Spectra and Lightcurve}

The brightest source in NGC 3379 (labeled ULX in Fig. 1) is located 
approximately $7^{\prime\prime}$ (360 pc)
to the NE of the AGN.   This source is not in a known
globular cluster, but it is difficult to detect globulars so close to the galactic center.  
We extracted a spectrum of this source using the region generated by {\it wavdetect} which
is approximately the 90\% encircled energy aperature at 1~keV.
A background spectrum was acquired from within the central $10^{\prime\prime}$ 
region of the galaxy after excising all emission from detected point sources.  
We first fitted the 0.3 to 3.0~keV spectrum with an absorbed multi-color disk (MCD)
blackbody model, but the fit was unacceptable ($\chi^2$/DOF =38/28). We then tried an 
absorbed power-law model and obtained an acceptable fit
($\chi^2$/DOF =27.8/28) with $\Gamma=1.6 \pm 0.3$ (90\% confidence 

\begin{inlinefigure}
  \center{\includegraphics*[angle=-90,width=1.00\linewidth,bb=40 40 572 709,clip]{f7.eps}}
  \caption{ACIS-S3 spectrum of the ULX along with the best-fit absorbed power-law model.}
\end{inlinefigure}

\bigskip

\noindent
limit) and galactic absorption (see Fig. 7). 
Adding a disk component to the power-law
model did not improve the fit further. The unabsorbed 0.3-10.0~keV flux 
is $2.0 \times 10^{-13}$~ergs~cm$^{-2}$~s$^{-1}$, which corresponds to 
L(0.3-10.0 keV)=~$2.7 \times 10^{39}$~ergs~s$^{-1}$.
The probability of detecting a background source with a flux equal to, or greater 
than this value within the central $7^{\prime\prime}$ of the galaxy is less 
than $10^{-4}$.

The intensity of this source varies smoothly by a factor of two during the
course of the 30 ksec observation (see Fig. 8).  
The peak luminosity of the source is $3.5 \times 10^{39}$~ergs~s$^{-1}$
which places it among the ULX class of objects and corresponds to the 
Eddington luminosity of a $30 \Mo$ black hole.
We also generated a hardness ratio light curve, but did not find significant
variations during the observation.

To check for long-term variability in the 
ULX, we examined the 24~ksec ROSAT HRI observation of NGC 3379 taken on Aug. 2, 1996 and 
found that the X-ray emission was centered on the ULX, not the center of the galaxy.
The net HRI count rate within a $10^{\prime\prime}$ radius aperture 
(corresponding to the 90\% encircled energy radius) centered on the
J2000 coordinates of the ULX as determined from the S3 image is $3.3 \times 10^{-3}$~ct~s$^{-1}$.
Assuming an absorbed power-law model with the best-fit parameters listed above 
gives an unabsorbed 0.3-10.0~keV flux of $3.0 \times 10^{-13}$~ergs~cm$^{-2}$~s$^{-1}$,
which is 50\% higher than the ACIS measured flux of the ULX in 2001. The HRI flux measurement
is undoubtedly affected by contamination from other point sources. Based on the S3 image,
the ULX contributes 60\% of the flux within the same apperature used to compute
the HRI flux.  Thus, there is strong evidence that the ULX was at a comparable 
luminosity 5 years prior to the Chandra observation.  There are insufficient counts in 
the HRI data to measure short term variability.

\section{Diffuse Emission and the AGN}

After excising the emission from all detected point sources,
we extracted a spectrum of the diffuse emission within the central $15^{\prime\prime}$ 
(770~pc).  A background spectrum was taken from a circular aperture 
$2.5^{\prime}$ from the center of the galaxy on the S3 chip.  The background
subtracted spectrum was then fitted to an absorbed thermal (MEKAL) plus power-law
model in the 0.3-5.0~keV band.  The best-fit power-law component has

\begin{inlinefigure}
  \center{\includegraphics*[width=0.90\linewidth,bb=19 152 570 700,clip]{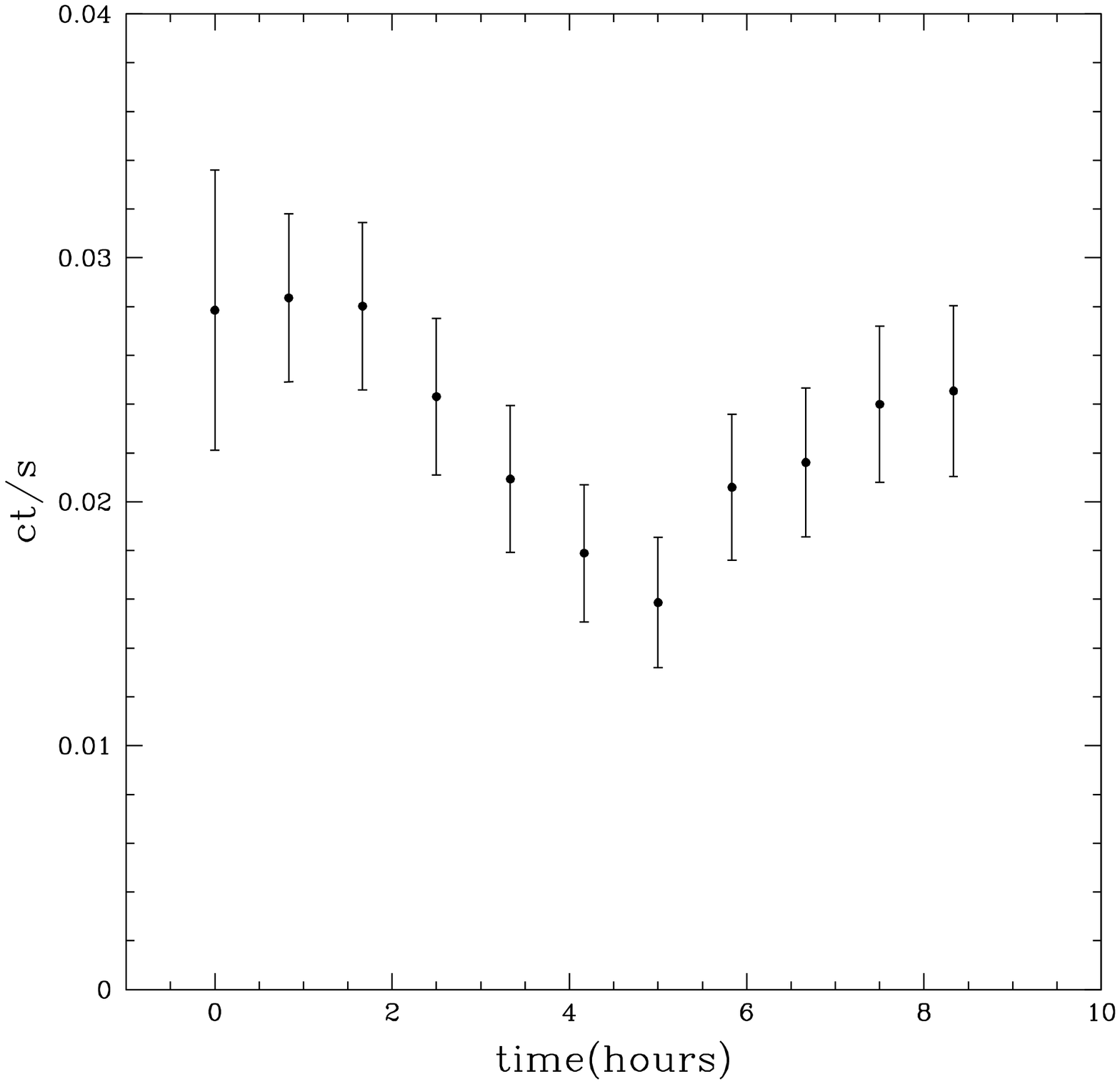}}
  \caption{The light curve of the ULX in the 0.3-6.0~keV band from the Chandra observation
  of 13 Feb, 2001.  The data are binned into 3~ksec intervals.}
\end{inlinefigure}

\bigskip

\noindent 
$\Gamma  = 1.7 \pm 0.3$ and an unabsorbed 0.3-10.0 keV flux of 
$5.5 \times 10^{-14}$~ergs~cm$^{-2}$~s$^{-1}$, corresponding to 
L(0.3-10.0 keV)=~$7.4 \times 10^{38}$~ergs~s$^{-1}$.  The best-fit thermal
component has kT=0.6 (0.4-1.0~keV) and an unabsorbed 0.3-10.0 keV flux of
$6.6 \times 10^{-15}$~ergs~cm$^{-2}$~s$^{-1}$, corresponding to
L(0.3-10.0 keV)=~$8.9 \times 10^{37}$~ergs~s$^{-1}$.  All errors are given
at the 90\% confidence limit.  This analysis shows that 90\% of the remaining
diffuse emission most likely arises from point sources with fluxes below the 
detection limit in the Chandra observation.  
Assuming the luminosity function derived above
is valid at lower luminosities, then the luminosity of the unresolved power-law 
component can be generated by approximately 40 sources below 
$3.0\times 10^{37}$~ergs~s$^{-1}$.

Using the best-fit emission measure in the thermal model and 
assuming a uniform gas distribution in the central 770~pc gives an electron
number density of $n_e = 0.01$~cm$^{-3}$, a gas mass of 
$5 \times 10^5 \Mo$, and a cooling time ($t_c = 5kTM_{gas}/2\mu m_p L$) of 
$8 \times 10^8$ yr.  If the gas is able to cool unimpeded, this gives
a mass cooling rate within the central 770~pc of $\dot {\rm M} = 6 \times 10^{-4} \Mo yr^{-1}$.
The estimated central gas density in NGC 3379 is a factor of 10 less than that found in 
X-ray luminous ellipticals whose emission is dominated by 
hot gas (e.g., Lowenstein et al. 2001).

The optical luminosity density in NGC3379 can be determined from the surface brightness profile
and Abel's formula (Binney \& Tremaine 1987).  Using the de Vaucouleurs profile for 
NGC 3379 found by Capaccioli et al. (1990), gives $L_B = 2.3 \times 10^{9} {\Lo}_B$ within 
the central 770~pc. Using the stellar mass loss rate from Faber \& Gallagher (1976)
for elliptical galaxies of $0.15\Mo/(10^{10} \Lo yr)$, gives a stellar mass loss
rate of $\dot {\rm M}_s = 0.035 \Mo$~yr$^{-1}$.  Thus, the 
presently observed gas mass in NGC 3379 can be produced by stars in only $10^7$ years, which
is significantly less than the cooling time, and only 5 times the sound crossing time in this region.
The continuous injection of gas shed by evolving stars with a temperature equal to the virial
temperature of the stellar system essentially negates the effects of radiative
cooling.  Such a small amount of gas implies that most of the mass shed by stars
has either been accreted by the central supermassive black hole or has been expelled from the central
region in a wind.

The point source labeled AGN in Fig. 1 is located within $1^{\prime\prime}$ 
of the optical centroid of the galaxy, and is probably associated with the central 
supermassive black hole with a dynamically measured mass of $1.0-3.9 \times 10^8 \Mo$
(Magorrian et al. 1998, Haring \& Rix 2004). 
The central black hole in NGC3379 is also a weak 1.4~GHz radio source with a luminosity
of $\nu L_{\nu} = 4.3 \times 10^{35}$ergs~s$^{-1}$ (Condon et al. 1998).
There are too few net counts in the S3 data to fit the spectrum
of the AGN, but assuming an absorbed power-law model with $\Gamma=1.6$ and 
galactic absorption gives L(0.3-10.0 keV)=$8.0 \times 10^{38}$ergs~s$^{-1}$, corresponding  
to $\sim 3 \times 10^{-8}$ of its Eddington limit.  
Using the density and temperature
of the ambient gas derived above gives an accretion radius for the central black
hole of $R_a = GM_{bh}/c_a^2 = 7$~pc, and a Bondi accretion rate 
$\dot {\rm M}_b = 4 \pi R_a^2 \rho_g c_a = 6 \times 10^{-5} \Mo yr^{-1}$,
which is a factor of 10 below the cooling rate within the central 770~kpc.  The implied 
efficiency, $\epsilon = L/\dot {\rm M} c^2$, of the central AGN 
assuming Bondi accretion is $2 \times 10^{-6}$.  This low efficiency is typical 
of that found for supermassive black holes 
at the centers of ellipticals (Lowenstein et al. 2001, DiMatteo et al. 2003).
The low luminosity of the central black holes in ellipticals can be 
explained by either a very low radiative efficiency, as in the advection dominated accretion 
flow model (ADAF; Narayan \& Yi 1994) or radiatively inefficient accretion flow
model (RIAF; Yuan \& Narayan 2005), or a reduction in the accretion rate,
as in models with both inflow and AGN driven outflows (e.g., Blandford \& Belgelman 1999).

While the uncertainty on 
the temperature of the diffuse gas is large, the best-fit temperature is 
twice the temperature associated with the velocity dispersion of the stars
(kT=$\mu m_p \sigma_p^2 /k=0.3$~keV for $\sigma_p = 217$~km~s$^{-1}$; Prugniel \& Simien 1996).
This suggests that the gas is being heated by the central AGN and is presently flowing out 
of the system in a galactic wind.  Assuming the gas is in a steady-state wind 
(i.e., $\dot M_s = 4 \pi r^2 \rho_g u_w$ at 770 pc) gives $u_w=20$~km~s$^{-1}$ and
an energy outflow rate (mostly thermal) of
$5 \times 10^{39}$ergs~s$^{-1}$, which is 4 orders of magnitude greater than
the radio luminosity of the AGN.  The low wind velocity near the galactic center is typical
of galactic wind models since the wind velocity continues to increase with increasing radius
(e.g., David, Forman \& Jones 1991).  The lack of detectable diffuse emission at
large radii in NGC 3379 may be due to the greater wind velocities and lower gas
densities.

Chandra has detected X-ray cavities filled with radio emitting plasma
and AGN driven shocks in many cluster cooling flows and individual elliptical galaxies
(e.g., McNamara et al. 2000, Finoguenov \& Jones 2002, Fabian et al. 2003, 
Blanton et al. 2003, Forman et al. 2005, Nulsen et al. 2005a, 
Nulsen et al. 2005b, McNamara et al. 2005). Based on the analysis of cavities found
in a sample of 16 clusters,
one group and one galaxy, Birzan et al. (2004) found that the ratio of mechanical
to radio power of the central AGN varies
from 10 in the most radio luminous AGNs, up to $10^4$ in systems with less radio 
luminous AGNs.  In clusters with AGN driven shocks, the shock energies are typically a few
times the energy in the cavities, which further increases the ratio of mechanical 
to radio power.  The largest shock and cavity energies yet observed are found
in MS0735.6+7421 (McNamara et al. 2005),
which has a ratio between mechanical and radio power of $10^5$.
Thus, at least on energetics grounds, the hypothesis that the lack of a significant reservoir 
of hot gas in NGC 3379 is due to an AGN driven wind is consistent with Chandra 
observations of systems perturbed by radio outbursts.  The low luminosities
of the central supermassive black holes in ellipticals may, in general, 
arise from the same feedback mechanism between the central AGN and hot gas 
as that observed in cluster cooling flows.

\section{The Nature of the ULX}

Based on ROSAT and Chandra observations, approximately 50\% of late-type galaxies 
and less that 10\% of early-type galaxies contain
ULXs (Ptak \& Colbert 2004, Irwin et al. 2004).
In late-type galaxies, the ULXs are primarily associated with regions 
of star formation, indicating that they are most likely black hole binaries (BHBs) 
with high-mass companions. However,  Colbert et al. (2004) recently estimated that 
20\% of the ULXs in spirals are not associated with recent star formation and 
could have low mass companions as is likely in early-type galaxies.
As noted in the introduction, ULXs could arise from non-isotropic, 
sub-Eddington emission from stellar mass black holes, super-Eddington 
emission from stellar mass black holes, or sub-Eddington emission from IMBHs.
van der Marel (2003) has noted that the X-ray luminosity function of 
X-ray point sources in galaxies can be fit by a single power-law, even to luminosities
above $10^{39}$~ergs~s$^{-1}$, suggesting that the accreting objects in ULXs are not a separate
class of objects (e.g., IMBHs). 
There are 18 BHBs in our Galaxy with dynamically measured black hole masses 
between $M_{bh}=3-18~\Mo$ (McClintock \& Remillard 2004).  
Assuming isotropic emission,
3 of these binaries are super-Eddington, with peak luminosities 
up to 7 times their Eddington-limit, suggesting that some of the extragalactic ULXs 
could be stellar mass black holes.  King (2003) has proposed that ULXs comprise
two separate classes of super-Eddington mass accretion rate systems.  
In regions of recent star formation in spiral galaxies, ULXs would arise from thermal-timescale
mass transfer in high mass X-ray binaries, while in elliptical galaxies,
the ULXs would be similar to the micro-quasars observed in our galaxy.
The long term stability of the ULX in NGC 3379 as determined from the ROSAT HRI and Chandra observations,
may pose a problem for the micro-quasar interpretation of ULXs in early-type galaxies.


The mass of the central compact object in a ULX can be estimated, if thermal
emission from an accretion disk can be detected, since the temperature at the 
inner edge of an accretion disk scales as $T_{in} \propto M_{bh}^{1/4}$.
Spectral analysis of ASCA data on ULXs indicated that these objects
could be described by the MCD model with inner disk temperatures
of $kT=1.1-1.8$~keV, implying stellar masses for the accreting objects
(Makishima et al. 2000). More recent analysis of Chandra and XMM-Newton
data, however, indicate that some ULXs can be described as pure power-laws
with $\Gamma=1.5-2.2$, while others require a power-law plus disk model 
(e.g., Roberts et al. 2001, Zezas et al. 2002a, Zezas et al. 2002b, 
Humphrey et al. 2003, Miller et al. 2004,
Liu, Bregman \& Seitzer 2002). Soft accretion disks have been detected
in two ULXs in NGC 1313 with inner disk temperatures of 150~eV, implying
masses of $100-1000 \Mo$  (Miller et al. 2003). While we cannot estimate the black hole
mass in the ULX in NGC 3379 due to the lack of detectable thermal emission 
from the disk, the best-fit power-law index ($\Gamma=1.7$) is consistent with recent 
Chandra and XMM-Newton results and similar to the low hard state observed in 
galactic BHBs (McClintock \& Remillard 2004)


The power-law spectrum of the ULX in NGC 3379 suggests that the emission is dominated by 
Compton up-scattering of soft disk photons in a optically thin corona.
The variability in the light curve could arise from a partial eclipse or 
absorption by cooler material in the outer parts of the accretion disk.
The slow rise and fall times in the light curve (see Fig. 8), along with the 
lack of any variation in hardness ratio, suggest that the intensity variation is 
due to a partial eclipse of the extended coronal emission with a 
period of 8-10 hours.  While variability on the scales of months to years is well known
for ULXs, there are only a few published cases of periodic variability
on the time scale of hours, all of which are in late-type galaxies 
(Sugiho et al. 2001, Circinus; Bauer 2001; M51 Liu et al. 2002;
NGC 628; Liu et al. 2005).  The ULX in NGC 3379 is the only known ULX in an early-type 
galaxy with possible periodic behavior in its light curve on the time scale of hours.

Orbital periods of 8-10 hr are very common for LMXBs (Verbunt 1993). Assuming that the secondary
star is filling its Roche lobe and transferring mass to the primary, we
can estimate the mass of the secondary.  Using the expression for the 
Roche lobe radius from Paczynski (1967) and Kepler's law gives
$P=8.9(R_2/R_{\odot})(\Mo / M_2)$~hr (Verbunt 1993).  For main sequence stars
$(R_2/R_{\odot}) = (M_2/ \Mo)$, indicating that if the period is 8-10 hr, then
the secondary star is approximately a solar mass.  The mass-radius relation 
for a He core burning star or a white dwarf predicts a much larger
mass for the secondary which is unlikely in an early-type galaxy.

\section{Summary}

The Chandra observation of the intermediate luminosity elliptical
galaxy NGC 3379 shows that only a small fraction of the gas shed
by evolving stars still resides within the hot ISM. A wavelet
detection algorithm resolves 75\% of the emission within the central
5~kpc into point sources.  The luminosity function of the point
sources detected at greater than $4 \sigma$ significance is consistent
with that found for other ellipticals observed by Chandra (Kim \& Fabbiano 2004).
Unlike other ellipticals observed by Chandra, 
only 8\% of the point sources are associated with globular clusters, which 
is comparable to the fraction of LMXBs in our galaxy.
The low specific frequency of globular clusters
and the low fraction of X-ray point sources associated with
globulars clusters in NGC3379 is actually more similar to Chandra observations
of S0 galaxies rather than ellipticals (e.g., Blanton, Sarazin \& Irwin 2001).

Spectral analysis of the unresolved emission 
within the central $15^{\prime\prime}$ (770~pc) indicates that 90\% of the
emission probably arises from point sources with fluxes below the
detection limit in the Chandra observation.  
If the luminosity function of the detected point sources is valid at 
lower luminosities, then the diffuse power-law emission can be accounted for 
by approximately 40 sources with luminosities below 
$3.0 \times 10^{37}$~ergs~s$^{-1}$.
The remaining 10\% of the unresolved emission from the central 770~pc is well described
by thermal emission with $kT=0.6$~keV.  Assuming a uniform gas density in this
region gives a gas mass of $5 \times 10^5 \Mo$, which can be supplied by 
stellar mass loss in $10^7$~years. Such a small amount 
of gas indicates that the stellar mass loss
is either being expelled from the central regions in a wind or is being 
accreted by the central black hole.  
The X-ray luminosity of the central AGN is $8 \times 10^{38}$~ergs~s$^{-1}$.
If gas is being accreted by the central black hole at either the Bondi accretion rate or mass 
cooling rate, then the radiative efficiency of the black hole must be $~\sim 10^{-6}$,
as in the ADAF or RIAF models models (Narayan \& Yi 1994 and Yuan \& Narayan 1995).
If the gas is flowing out of the system in a wind, the 
energy outflow rate would be $5 \times 10^{39}$~ergs~s$^{-1}$, which is 
4 orders of magnitude greater than the 1.4~GHz radio power of the AGN. 
Such a large ratio between AGN mechanical and radio power is commonly found among
cluster cooling flows with X-ray cavities and shocks 
(Birzan et al. 2004; Nulsen et al. 2005a; Nulsen et al. 2005b; McNamara et al. 2005).

The most luminous source in NGC3379 is located 360~pc from the central AGN with 
a peak luminosity of $3.5 \times 10^{39}$ergs~s$^{-1}$, corresponding to the
Eddington luminosity of a $30 \Mo$ object.  The spectrum of this source
is well fitted with an absorbed power-law model with $\Gamma=1.7$, which is 
similar to other ULXs observed by Chandra and XMM-Newton and
the low-hard state of galactic black hole binaries.  Examining the archival
ROSAT HRI observation of NGC 3379 shows that the ULX was at a comparable
luminosity 5 years prior to the Chandra observation.  
The long term stability of the ULX 
may pose a problem for the micro-quasar interpretation of ULXs in early-type galaxies.

The light curve of the ULX in NGC 3379 varies smoothly by a factor of two during the 
Chandra observation. The slow rise and fall times in the light curve
and the consistency of the power-law spectrum during the observation all
suggest that the ULX is undergoing a partial eclipse of the 
extended corona surrounding an accretion disk with a period of 
8-10~hr.  
Assuming the secondary is a main sequence star filling its Roche lobe 
gives a mass for the secondary of approximately $1 \Mo$.  Variability has 
been observed in other ULXs, but the ULX in NGC 3379 is the only ULX in an 
elliptical galaxy with possible periodic behavior.
Due to the high surface brightness density of sources in the central 1~kpc of
NGC 3379, only a long Chandra observation can determine if the lightcurve of 
the ULX is truly periodic.

\end{document}